%% file: JFM-template.tex
\documentclass[]{jfm}
\usepackage{graphicx}
\usepackage{newtxtext}
\usepackage{newtxmath}
\usepackage{natbib}
\usepackage{hyperref}
\usepackage{multirow}
\usepackage{textcomp}
\usepackage[justification=justified,singlelinecheck=false]{caption}
\usepackage{subcaption}
\usepackage{etoolbox}
\usepackage{rotating}
\usepackage{pifont}
\hypersetup{
    colorlinks = true,
    urlcolor   = blue,
    citecolor  = black,
}
\newcommand{\RomanNumeralCaps}[1]
\linenumbers

\title[Mean velocity in stably stratified turbulent channel flow]{Mean velocity profile in stably stratified {turbulent channel flow}}

\author{Sanath Kotturshettar\aff{1}
  \corresp{\email{s.b.kotturshettar@tudelft.nl}},
  Pedro Costa\aff{1}
 \and Rene Pecnik\aff{1}}

\shortauthor{S.~Kotturshettar, P.~Costa and R.~Pecnik}

\affiliation{\aff{1}Process \& Energy Department, TU Delft, Leeghwaterstraat 39, 2628 CB, Delft, The Netherlands.}

\begin{document}
\maketitle

\begin{abstract}
The Monin–Obukhov Similarity Theory (MOST) is a cornerstone of atmospheric science for describing turbulence in stable boundary layers. Extending MOST to stably stratified turbulent channel flows, however, is non-trivial due to confinement by solid walls. In this study, we investigate the applicability of MOST in closed channels and identify where and to what extent the theory remains valid. 
A key finding is that the ratio of the half-channel height to the Obukhov length serves as a governing parameter for identifying distinct flow regions and determining their corresponding mean velocity scaling.
Hence, we propose a relation to estimate this ratio directly from the governing input parameters: friction Reynolds and friction Richardson numbers ($Re_{\tau}$ and $Ri_{\tau}$). The framework is tested against a series of direct numerical simulations (DNS) across a range of $Re_{\tau}$ and  $Ri_{\tau}$. 
The reconstructed velocity profiles enable accurate prediction of the skin friction coefficient crucial for quantifying pressure losses in stratified flows in engineering applications. 
\end{abstract}

\section{Introduction}
\label{sec:introduction}

\input{intro_new_new}

{\section{Suitability of MOST for pressure-driven channels}}
\label{sec:most_channel}

\input{most_channels}

\input{result_revision}

\section{Conclusion}
\label{sec:conclusion}

\input{conclusion}
\bibliographystyle{jfm}
\bibliography{jfm_cleaned}
\newpage
\appendix

\input{appendix}

\end{document}

%% file: intro_new_new.tex
Stratified turbulent flows are prevalent in environmental and engineering systems. Examples include atmospheric currents that affect pollutant dispersion, nutrient transport in oceans, and fluid flows in energy conversion systems. In case of stably stratified flows, where a lighter fluid overlies a heavier fluid, gravity acts to preserve the stratification, while turbulence promotes mixing of layers. The resulting flow behavior is governed locally by the relative dominance of these competing mechanisms \citep{brethouwer_scaling_2007, zonta_stably_2018}. Accurately predicting these flows is crucial in weather forecasting and determining key design parameters for engineering applications, such as the friction factor and the Nusselt number.

\begin{figure}
     \centering
     \includegraphics[width=1\linewidth]{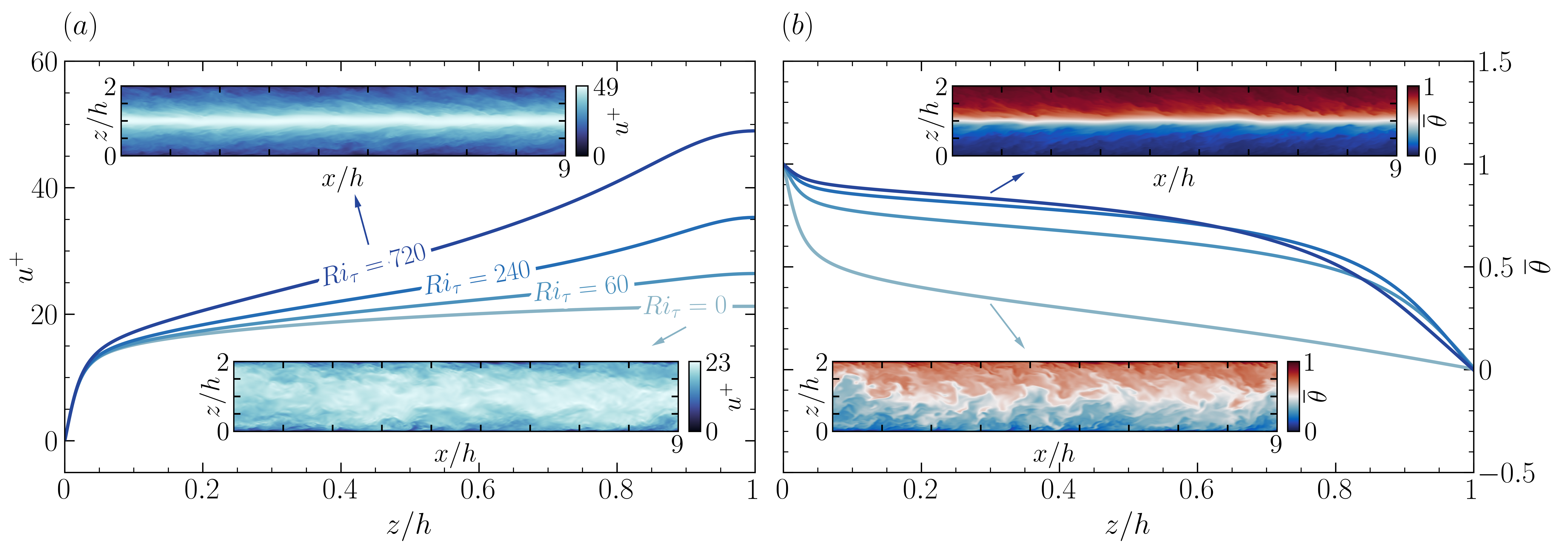}
    \caption{\emph{(a)} Mean velocity profiles and \emph{(b)} Mean temperature profiles for friction Richardson numbers $Ri_{\tau} = \Delta \rho g h/\rho_0 u_{\tau}^2 = 0, \ 60,\ 240,\ 720$, represented with increasing darkness. Instantaneous contours at the mid-spanwise plane illustrate flow structures for neutral ($Ri_{\tau} = 0$)  and stratified ($Ri_{\tau} =  720$) cases. All profiles and contours correspond to a friction Reynolds number $Re_{\tau} = \rho_0 u_{\tau} h/\mu_0 = 550$. In these governing parameters, $u_{\tau}$ is the friction velocity, $\rho_0$ and $\mu_0$ denote the density and dynamic viscosity of the fluid, $h$ is the half-channel height, $g$ is the gravitational acceleration (pointing in the negative $z$ direction), and $\Delta \rho$ is the imposed density difference across walls.}
    \label{fig:inst_cont_vel}
\end{figure}

In wall-bounded turbulent flows, the presence of stable stratification leads to a reduction in both the skin friction coefficient and the Nusselt number, as shown by \citet{fukui_laboratory_1983}. This reduction arises from the suppression of turbulent momentum and heat transport by buoyancy forces, especially away from the wall where shear is weak \citep{garg2000stably, armenio2002investigation}. Figure~\ref{fig:inst_cont_vel} illustrates this in instantaneous streamwise velocity and temperature contours for stratified and neutral (unstratified) channel flows. While velocity/temperature variations associated with near-wall flow structures can be observed in both cases, turbulent fluctuations can be severely damped in the outer region of the stratified channel due to buoyancy. As a result, the mean velocity and temperature gradients steepen near the channel center; see the profiles near $z/h=1$ in figure~\ref{fig:inst_cont_vel}. Indeed, density gradients in the bulk can give rise to laminar internal gravity waves (IGWs) that interact with the turbulent regions of the flow \citep{iida2002direct}, as visible in the contours of the stratified case in figure~\ref{fig:inst_cont_vel}. {Buoyancy effects suppress turbulence over a larger fraction of the channel compared to the ABL, since turbulent shear linearly decreases from the wall, even leading to flow laminarization at the centerline \citep{moestam2005numerical,armenio2002investigation,garcia2011turbulence}.}
The underlying mechanism for the suppression of turbulent fluxes is the restriction of turbulent kinetic energy (TKE) redistribution across the channel height. Here, vertical turbulent mixing against a gravitational potential gradient results in the conversion of TKE into potential energy. This energy exchange can result in irreversible mixing of the density field, leading to TKE dissipation \citep{caulfield_layering_2021}.  
\par
Stably stratified wall-bounded turbulent flows have been extensively studied in the context of stable atmospheric boundary layers (ABL) to describe the modulation of turbulent fluxes near the land surface. The most established framework used to predict the mean flow in this context is the Monin--Obukhov Similarity Theory (MOST), based on the seminal work of \cite{monin1954basic}. 
In MOST, mean velocity gradients $({\mathrm{d}\overline{u}}/{\mathrm{d}z})$ in the overlap region of a fully developed turbulent boundary layer subjected to stratification are related to the wall-normal distance $(z)$ and the Obukhov length scale $(L)$ \citep{obukhov1971turbulence} as 
\begin{equation}\label{eq:obukhovL}
    \frac{\kappa z}{u_{\tau}} \frac{\mathrm{d}\overline{u}}{\mathrm{d}z} = \phi\left(\frac{z}{L}\right);
    \quad \text{with}\quad L = -\frac{u_{\tau}^{3}}{\kappa\ ({g}/{\theta_0})\ q_{\mathrm{w}}},
\end{equation}
where $q_w$ is the surface heat flux, $g/\theta_0$ is the buoyancy parameter, and $\kappa$ is the von Kármán constant. 
In this context, $L$ estimates the height above which buoyancy starts to quench turbulent mixing, because buoyancy destruction of turbulence overwhelms shear production. The similarity function $\phi$ encapsulates the effect of stratification and is typically determined from atmospheric field data. 

Further on, \citet{nieuwstadt1984turbulent} reasoned that the dynamics of stratified turbulence are better characterized by \emph{local} rather than surface quantities, and introduced a scale based on local momentum and heat fluxes, $\overline \tau(z)$ and $\overline q(z)$. This results in a local Obukhov length scale, \begin{equation}\label{eq:local_obukhov_length}
     \Lambda(z) = -\frac{\left(\overline \tau(z)/ \rho_0\right)^{3/2}}{\kappa\ ({g}/{\theta_0})\ \overline q(z)},
\end{equation}
that extended the validity of MOST from the surface layer to the entire boundary layer \citep{nieuwstadt1984turbulent, holtslag1986scaling,sorbjan1986similarity}. Though several refinements have been proposed over the years \citep{salesky_buoyancy_2013, grachev_similarity_2015, li_connections_2016, stiperski_generalizing_2023}, the core principles of MOST remain remarkably valid. In striking contrast, its validity for describing internal wall flows remains largely unexplored. %
 
Several detailed numerical studies have investigated the interaction between turbulence and stratification \citep[see][for a review]{zonta_stably_2018}. \citet{nieuwstadt2005direct} performed the first DNS of open-channel flow driven by a constant pressure gradient, with stable stratification maintained by extracting heat from the lower wall. Building on these simulations, \citet{van_de_wiel_cessation_2012} and \citet{donda_collapse_2015} demonstrated that MOST can be applied to derive mean velocity profiles in open channel flows.
Differentially heated, closed turbulent channel flows with stable stratification have been studied since the early works by \citet{garg2000stably} and \citet{armenio2002investigation}. Here, as we have seen, strong density gradients form near the channel center, promoting the formation of IGWs \citep{iida2002direct}. \citet{garcia2011turbulence} provided a detailed spectral analysis of these flows, demonstrating that the outer layer flow structures scale with $\Lambda$.  More recently, \citet{zonta2022interaction} proposed empirical correlations for predicting the friction coefficient and Nusselt number in stably stratified channel flows. Despite these advances, the applicability of MOST to describe the mean flow has not been explored.\par

The goal of this study is to address this gap and the following key question: `\emph{How can MOST be extended to describe the mean flow of a stably-stratified turbulent channel?}' %
In this work, we evaluate the validity of MOST in closed channels to trace the full mean velocity profile. To this end, we characterize the different flow regions considering the dominant balance between buoyancy and shear effects, and applying the corresponding local scalings. This involves not only deriving velocity scaling laws in the different regions, but also delineating the corresponding boundaries. Finally, we integrate the reconstructed mean velocity profile to estimate the skin friction coefficient in stably stratified channel flow.

\section{Computational campaign}

We perform direct numerical simulations (DNS) of stably stratified turbulent channel flows to develop a framework for describing their mean velocity profile, which is subsequently validated using the simulation data. The Navier–Stokes equations are solved in the low-Mach number regime \citep{MAJDA1985, COOK1996263}, with a prescribed temperature difference of $1\%$. This setup ensures minimal variation in thermodynamic properties, enabling a reasonable comparison with cases in the literature that use the Oberbeck–Boussinesq approximation \citep[see, e.g.,][]{garcia2011turbulence}. The governing equations are:
\begin{equation}
    \frac{\partial \rho}{\partial t} + \frac{\partial \rho u_j}{\partial x_j} = 0,
    \label{eq:con}
\end{equation}
\begin{equation}\label{eq:mom}
    \rho \frac{\partial u_i}{\partial t} + \rho u_j\frac{\partial u_i}{\partial x_j} = -\frac{\partial p}{\partial x_i} + \frac{\partial \tau_{ij}}{\partial x_j}+\rho g_i+f_i\delta_{i,1},
\end{equation}
\begin{equation}\label{eq:ene}
    \rho C_p\frac{\partial T}{\partial t} + \rho C_p u_j\frac{\partial T}{\partial x_j} = \frac{\partial}{\partial x_j}\left(\lambda \frac{\partial T} {\partial x_j}\right)  + \frac{\mathrm{d} p_0}{\mathrm{d} t},
\end{equation}
where, $\tau_{ij} = \mu (\partial u_i/ \partial x_j + \partial u_j/ \partial x_i - 2/3\ \partial u_k/\partial x_k\ \delta_{i,j})$. Here, $p_0$ denotes the spatially invariant thermodynamic pressure, while $p$ represents the hydrodynamic pressure. The walls are maintained at fixed temperatures to enforce stable stratification. The arithmetic mean of the wall temperatures is chosen as the reference temperature at which the reference thermodynamic properties such as thermal conductivity ($\lambda$), heat capacity ($C_p$), dynamic viscosity ($\mu$), and density ($\rho$) are evaluated. The density is related to temperature via the ideal gas law ($\rho = p_0/ R T$), where $R$ is the universal gas constant. The gravitational acceleration vector $g_i$ acts vertically downward, and $f_1$ represents the constant pressure difference driving the flow along the streamwise direction. Periodic boundary conditions are imposed in the streamwise and spanwise directions, while the no-slip condition is enforced at the walls. The equations were solved on staggered Cartesian grids, with velocities at cell faces and scalars at the cell centers. The convective and diffusive terms were discretized in space using a second-order, finite volume scheme \citep{COSTA20181853}, and advanced in time using Wray's low-storage third-order Runge-Kutta method, with implicit treatment for wall-normal diffusion.\par

\begin{figure}
    \centering
    \includegraphics[width=\linewidth]{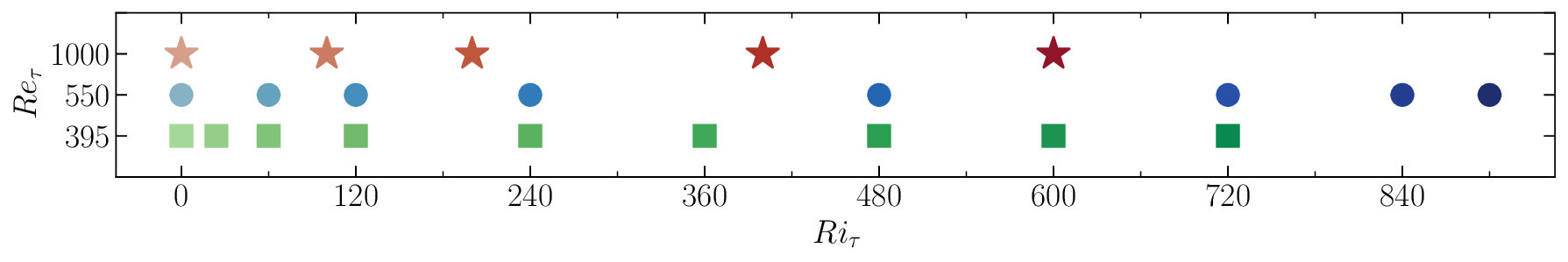}
    \caption{Summary of the parameter space sampled in the DNS database, with data at $Re_{\tau} = 1000$ from \citet{zonta2022interaction}. Data for each case will be presented consistently with these marker colors.}
    \label{fig:cases}
\end{figure}

This setup is characterized by a friction Reynolds number $Re_{\tau} = \rho_0 u_{\tau} h/\mu_0$, and a friction Richardson number, $Ri_{\tau} = \Delta \rho g h/\rho_0 u_{\tau}^2$ \citep{garg2000stably}; $\rho_0$ and $\mu_0$ denote the fluid density and dynamic viscosity at a reference temperature $\theta_0$, $h$ is the half-channel height, and $\Delta \rho$ the bottom-to-top wall density difference. The DNS were performed at $Re_{\tau} = 395$ and $550$. Additionally, the data at $Re_{\tau} = 1000$ was obtained from \citet{zonta2022interaction}. For each $Re_{\tau}$, a range of $Ri_{\tau}$ was explored ($0 \leq Ri_{\tau} \leq 900$), at fixed Prandtl number $Pr = 0.71$. Figure~\ref{fig:cases} summarizes the cases.

All simulations were conducted in a domain of dimensions, $L_x \times L_y \times L_z = 6\pi h \times 2\pi h \times 2h$, in the streamwise, spanwise, and wall-normal directions, respectively. At $Re_{\tau} = 395$, nine different $Ri_{\tau}$ were simulated using a grid of $ N_x \times N_y \times N_z = 1024 \times 512 \times 300$ points. On the other hand, eight different $Ri_{\tau}$ were simulated at $Re_{\tau} = 550$ using a grid of $ N_x \times N_y \times N_z = 1536 \times 768 \times 480$ points. Uniform grid spacing was used in the periodic directions, while wall-normal grid stretching was applied to resolve the Kolmogorov length scale near the wall. Previous studies (e.g., \citet{garcia2011turbulence}) have indicated that increasing stratification does not affect the smallest turbulent scales, so grid resolution requirements remain unchanged with increasing $Ri_{\tau}$. Simulations at higher $Ri_{\tau}$ were initialized using steady-state fields from the preceding lower $Ri_{\tau}$ case, following the strategy employed in earlier studies. Grid details of cases at $Re_{\tau} = 1000$ are available in \citet{zonta2022interaction}.

%% file: most_channels.tex
{
While MOST was originally formulated for the ABL, we investigate its suitability in capturing the effects of stratification in turbulent channel flow. The applicability is assessed by examining the dependence of normalized velocity gradients on the wall-based $(z/L)$ and local $(z/\Lambda)$ stability parameters, as introduced in Eqs.~(\ref{eq:obukhovL}--\ref {eq:local_obukhov_length}).  %
}

{
The function $\phi$, commonly referred to as the stability correction function, is typically obtained empirically by fitting experimental data. The linear relations proposed by \cite{webb1970profile, businger1971introduction, dyer1974review} provide a good fit for observations within the dry atmospheric boundary layer, particularly under weak stratification. However, the functional form is not universal, with these studies reporting slightly different slopes for the linear relation, typically ranging between $4.5$ and $5.5$.
}
{
Moreover, several alternative formulations for $\phi$ have also been proposed, including higher-order functions that reduce to a linear form under weak stratification but approach a constant value under strong stratification \citep{beljaars1991flux, chenge2005flux}. Through their linear behaviour under weak and non-linear corrections under strong stratification, these formulations recover the well-known `$z$-less' region observed in the ABL \citep{nieuwstadt1984turbulent}. However, this region has not been observed in channel flows \citep{armenio2002investigation, garcia2011turbulence, zonta_stably_2018}.}

\begin{figure}
    \centering
    \includegraphics[width= .8\linewidth]{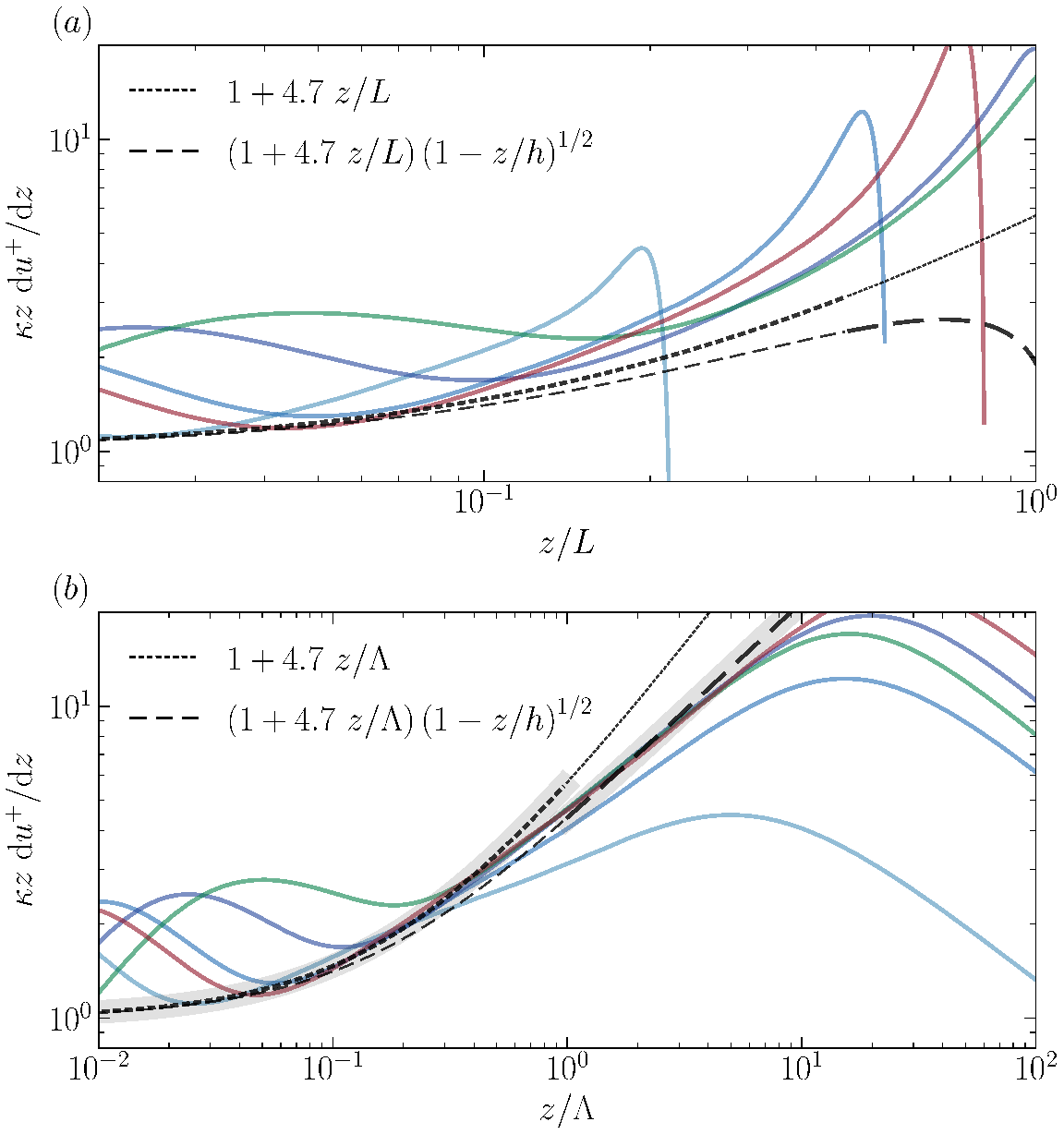}
    \caption{The dimensionless velocity gradients, $(\kappa z\ \mathrm{d} u^+/\mathrm{d}z)$, obtained from DNS across a range of stratification levels as a function of $z/L$ in panel \emph{(a.)} and $z/\Lambda$ in panel \emph{(b.)} Increasing shades of blue correspond to $Ri_{\tau} = 60,\ 240,\ 720$ at $Re_{\tau} = 550$, the green line corresponds to $Ri_{\tau} = 720$ at $Re_{\tau} = 395$, and the red one to $Ri_{\tau} = 600$ at $Re_{\tau} = 1000$. These gradients are compared against the Businger relation,  $1+4.7\ \zeta$, where $\zeta$ is the stability parameter.}
    \label{fig:phi_zeta}
\end{figure}

{
In turbulent channel flows, the mean velocity (and, thus, its gradient) has been reported to increase with stratification. This is attributed to the reduction in turbulent shear stress under stable stratification. To represent this behavior, we adopt the linear formulation proposed by \citet{businger1971introduction}, which captures the monotonic increase in velocity gradients with increasing stability.
}
{
Figure~\ref{fig:phi_zeta} presents the dimensionless velocity gradients, $(\kappa z/u_{\tau} \ \mathrm{d} u/\mathrm{d}z)$, obtained from DNS across a range of stratification levels as a function of the stability parameter: $z/L$ in panel~(a), and $z/\Lambda$ in panel~(b).
The figure also shows the Businger relation $1 + 4.7\zeta$, where $\zeta$ is the stability parameter. Clearly, the local scaling captures the variation in velocity gradients more consistently than the wall-based scaling, particularly in representing the increase in gradient magnitude.} 
{Interestingly, although the Businger relation has been primarily validated within the `constant-flux' layer, we find that it captures the velocity gradients reasonably well as long as turbulent shear remains the dominant mechanism and stratification effects are relatively weak ($z/\Lambda < 1$). For $z/\Lambda > 1$, deviations from the linear relation become evident, with the Businger relation tending to overestimate the velocity gradients (see short dash-dotted lines in figure~\ref{fig:phi_zeta}). 
}

{
To improve the similarity hypothesis, in the context of ABL, \citet{yokoyama1979vertical} proposed using `$z$'-dependent surface parameters (e.g., $u_{\tau}(z) = u_{\tau}(1 - z/h)^{\gamma}$) that reliably estimate local fluxes. This method was subsequently adopted by \citet{sorbjan1986similarity} and more recently by \citet{gryning2007extension} and \citet{Shen_Liu_Lu_Stevens_2025}. The exponent $\gamma$ is determined empirically and varies across different studies. Following a similar approach, we apply a correction of the form $(1 - z/h)^{\gamma}$ to the Businger relation, with $\gamma = 1/2$. As shown in figure~\ref{fig:phi_zeta}(b), the corrected relation captures the trend of the velocity gradients reasonably well for $z/\Lambda > 1$, particularly in high $Ri_{\tau}$ cases, where this region is more distinct. Furthermore, the value of $\gamma=1/2$ also aids analytical derivation through simplification of equations which will be discussed in the following sections.}

{
In the region where $z/\Lambda  > 1$, note that the current approach is equivalent to using a local friction velocity, $u_{\tau}(z) = u_{\tau}(1 - z/h)^{1/2}$, to normalize the velocity in Eq.~\eqref{eq:obukhovL} $(\kappa z\ \mathrm{d} u^+/\mathrm{d}z)$, while maintaining the linear Businger relation. However, for the remainder of the article, we interpret this factor as a correction to the Businger relation rather than as an estimation of the local friction velocity, and use a mixing length model that consistently accounts for these variations.
}

%% file: result_revision.tex
\section{Heat and momentum flux balance}

Following \citet{donda_collapse_2015}, $\Lambda$ can be expressed in terms of the wall-normal coordinate using the heat and momentum flux distributions. For pressure-driven turbulent channel flows, with walls at fixed temperatures, the mean total heat flux ($q$) is constant across the channel, while the total shear stress ($\tau$) varies linearly:
\begin{subequations}
\begin{align}
    q(z) &=  {\alpha}\frac{\mathrm{d}\overline{\theta}}{\mathrm{d}z} - \ \ \ \ \overline{w^{\prime}\theta^{\prime}} = q_{\mathrm{w}},\label{eq:heat_balance}\\
    \tau(z) &= \mu \frac{\mathrm{d} \overline{u}}{\mathrm{d} z} - {\rho_0}{\overline{u^{\prime}w^{\prime}}} = {\tau_{\mathrm{w}}} \left(1 - \frac{z}{h}\right),\label{eq:stressbalance}
\end{align}
\end{subequations}
where $\alpha$ is the thermal diffusivity, and the subscript `$\mathrm{w}$' denotes wall values. Substituting Eqs.~\eqref{eq:heat_balance} and \eqref{eq:stressbalance} into the definition of  $\Lambda$ (Eq.~\eqref{eq:local_obukhov_length}) yields:
\begin{equation}\label{eq:ob_lo_ob}
    \Lambda(z) = L \left(1 - \frac{z}{h}\right)^{3/2}, %
\end{equation}
with $L$ the Obukhov length scale as given in Eq.~\eqref{eq:obukhovL}.
Approximating the turbulent flux in Eq.~\eqref{eq:stressbalance} using the mixing length, we can write the stress balance in the well-known form %
\begin{equation}\label{eq:nu_t_stress_balance}
    \nu \frac{\mathrm{d} \overline{u}}{\mathrm{d} z} + \ell_m^2 \left|\frac{\mathrm{d}\overline{u}}{\mathrm{d}z}\right| \frac{\mathrm{d} \overline{u}}{\mathrm{d} z} = u_{\tau}^2 \left(1 - \frac{z}{h}\right),
\end{equation}
where $u_\tau = \sqrt{\tau_w/\rho_0}$ and $\nu = \mu/\rho_0$. This classical zero-equation model forms the basis for the proposed scaling laws. The mean velocity profile is naturally obtained by integrating Eq.~\eqref{eq:nu_t_stress_balance} across the channel.

{
The mixing length $(\ell_m)$ is governed by the relative influence of buoyancy forces and turbulent shear. It is defined by scaling its value for pressure-driven turbulent channel flows in neutral limit $(\ell_m^{\mathrm{N}})$, with the stability function $\phi$ that accounts for effects of stable stratification,
\begin{equation}\label{eq:ml_str}
\ell_m = \frac{\ell_m^{\mathrm{N}}}{\phi},
\end{equation}
with $\phi$ being a function of $z/\Lambda$ (recall the discussion in \S\ref{sec:most_channel}). For $\ell_m^{\mathrm{N}}$, we adopt the mixing length reported by \citet{pirozzoli_revisiting_2014} for channel flows,
\begin{equation}\label{eq:ml_neutral}
\ell_m^{\mathrm{N}} = \kappa z\left(1 - \frac{z}{h}\right)^{1/2}.
\end{equation}
{
Note that, in Eq.~\eqref{eq:ml_str}, $\ell_m^{\mathrm{N}}$ does not depend on the strength of stratification, and Eq.~\eqref{eq:ml_neutral} is used across the entire channel. In contrast, the function $\phi$ accounts for stratification, and the stability correction should be used depending on the value of $z/\Lambda$, as discussed in \S\ref{sec:most_channel}.
}
Hence, based on the local balance between viscous and turbulent stresses, and the competing influence of buoyancy and shear, Eq.~\eqref{eq:nu_t_stress_balance} can be simplified locally to identify distinct flow regions, as discussed in the next section.
}
\section{Proposed mean velocity scaling}
\begin{figure}
\centering
    \includegraphics[width = \linewidth]{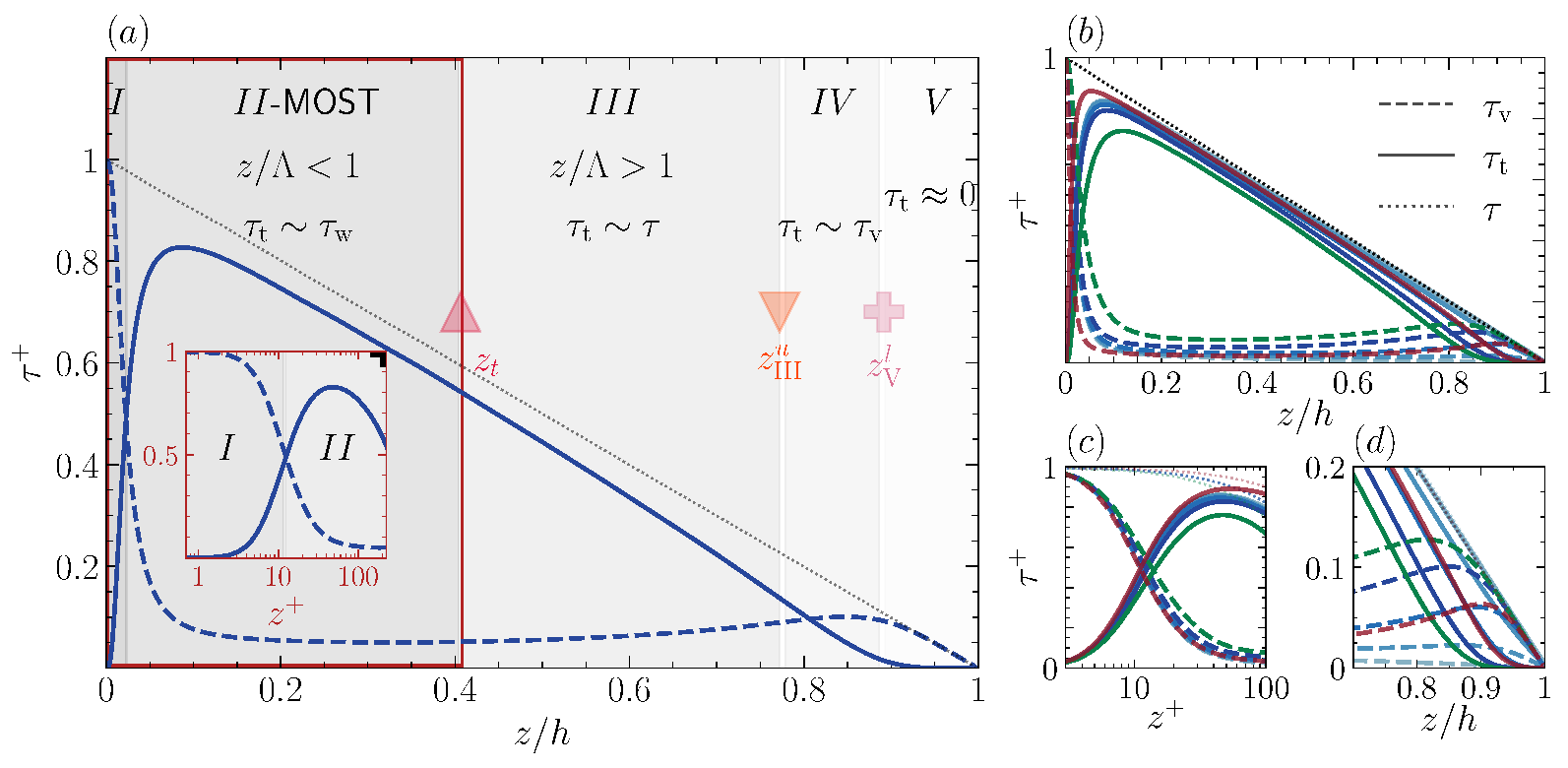}
    \caption{\emph{(a)} Schematic illustrating the distinct regions in a stably stratified turbulent channel flow, classified based on the influence of stratification (quantified by $z/\Lambda$) and the relative contributions of viscous stress ($\tau_{\mathrm{v}}$), turbulent stress ($\tau_{\mathrm{t}}$), to total stress ($\tau$). The identified regions are:  \emph{viscous sublayer (I), {shear-dominated sublayer} (II), {stratified outer region} (III), turbulent-viscous transition layer (IV)}, and \emph{viscous core (V)}. The illustration corresponds to the case with $Ri_{\tau} = 720$ at $Re_{\tau} = 550$. \emph{(b)} Stress profiles for various cases. Blue curves: $Re_{\tau} = 550$, $Ri_{\tau} = 0,\ 60,\ 240,\ 720$ (increasing darkness). Green: $Re_{\tau} = 395$, $Ri_{\tau} = 720$. Red: $Re_{\tau} = 1000$, $Ri_{\tau} = 600$ \citep{zonta2022interaction}. \emph(c) Close-up of the profiles close in the near wall region to illustrate the slow rise of turbulent stresses with increasing stratification. \emph{(d)} Close-up near the center of the channel, highlighting the drop-off of turbulent stresses to zero.}
    \label{fig:stress_schematic}
\end{figure}

Figure~\ref{fig:stress_schematic} shows the typical distribution of viscous and turbulent shear stresses in a pressure-driven turbulent channel flow. Our analysis highlights five characteristic regions as illustrated in figure~\ref{fig:stress_schematic}(a), based on the distribution of the local stability parameter ($z/\Lambda$) and the relative magnitudes of viscous and turbulent stresses. For convenience of reporting, the regions are hereafter given descriptive names.

Closest to the wall, the classical \emph{viscous sublayer (I)} (and the buffer layer) is unaffected by buoyancy forces as the strong shear dominates the stratification effects in this region. This is followed by an \emph{{shear-dominated sublayer} (II)}, where turbulent stresses dominate with minimal stratification effects. Further away from the wall, buoyancy effectively suppresses the larger eddies while the turbulent stresses are still much greater than the viscous stresses -- a region termed here as the \emph{{stratified outer region} (III)}. Beyond lies the \emph{turbulent-viscous transition layer (IV)}, where viscous and turbulent stresses become comparable. Finally, from the channel center, turbulence may vanish and lead to a laminar region, forming a \emph{viscous core layer (V)} governed by viscous stresses. We now derive the mean velocity profile in these regions, excluding the well-established $u^+ = z^+$ scaling in the viscous sublayer.%

\subsection{{Shear-dominated sublayer} (II) -- classical MOST}\label{sec:iner_sl}

Following the viscous sublayer, viscous stresses rapidly decrease while turbulent stresses increase, reaching a peak at approximately $z^+ \approx 30$ wall units (seen in figure~\ref{fig:stress_schematic}(c)). In the {shear-dominated sublayer}, the local stability parameter $z/\Lambda$ is less than one, indicating that turbulent shear dominates over buoyancy effects. 
This is typically classified as the `surface layer' in the ABL context, where the turbulent fluxes remain approximately constant and the classical MOST is applicable. In this region, Eq.~\eqref{eq:nu_t_stress_balance} simplifies to:
\begin{equation}\label{eq:reg_ine_sl}
    \ell_m \frac{\mathrm{d} \overline u}{\mathrm{d} z} = u_{\tau}\left(1 - \frac{z}{h}\right)^{1/2}.
\end{equation}

{
The mixing length in the shear-dominated sublayer $(II)$, denoted as $\ell_m^{\mathrm{II}}$, follows the formulation in Eq.~\eqref{eq:ml_str}. In this region, the effect of stable stratification is incorporated through the stability function  represented by the linear Businger relation, {$\phi = 1 + 4.7 \zeta$, where $\zeta = z/\Lambda$ corresponds to the local stability parameter (recall figure~\ref{fig:phi_zeta} and the discussion thereabout).} Hence, the mixing length is given by
\begin{equation}\label{eq:ml_iner_sl}
    \ell_m^{\mathrm{II}} = \frac{\ell_m^{\mathrm{N}}}{\phi} =\frac{\ell_m^{\mathrm{N}}}{1 + 4.7 \ \zeta} = \frac{\kappa z}{1 + 4.7 \ \zeta}\left(1 - \frac{z}{h}\right)^{1/2}\text{;}
    \quad \ \ \
    \zeta = \frac{z}{\Lambda}.
\end{equation}
}

{
Figure~\ref{fig:ml_compare} presents a comparison between the mixing length measured from DNS data, $(-\overline{u'w'})^{1/2}/(\mathrm{d}\overline{u}/\mathrm{d}z)$, and the expression given by Eq.~\eqref{eq:ml_iner_sl}. A good agreement is observed in the region where turbulent shear dominates (i.e., $z/\Lambda < 1$, see the fine dashed lines). The expression begins to deviate from the DNS profile when buoyancy effects become prominent, indicating a transition to a different regime where an alternative mixing length formulation is required.
\begin{figure}
    \centering
    \includegraphics[width=.75\linewidth]{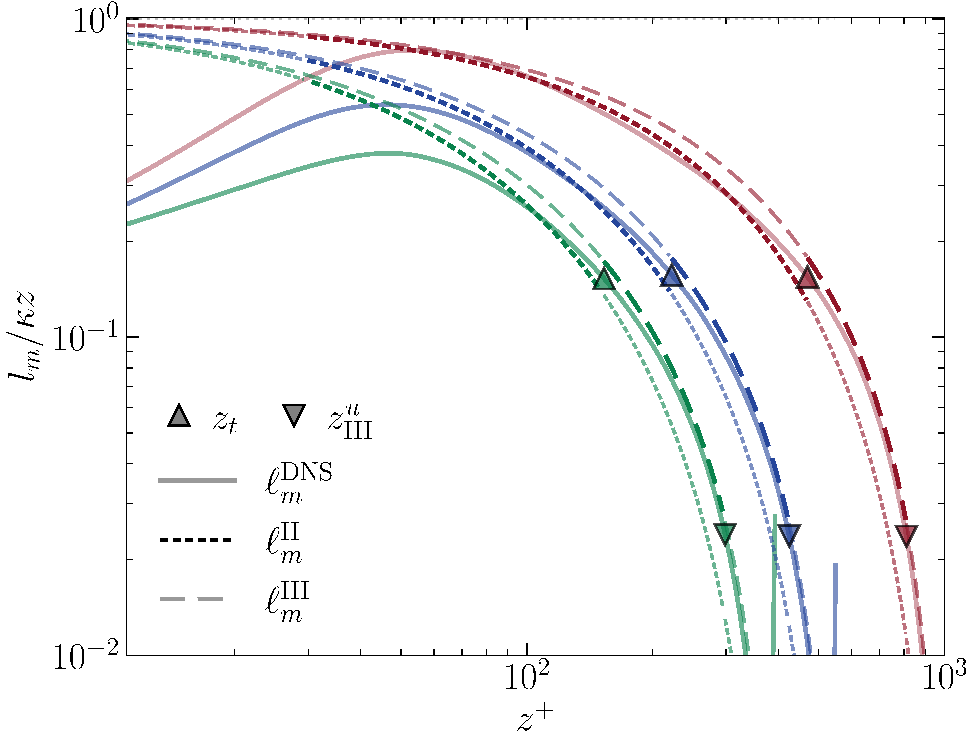}
    \caption{Comparison of the mixing length obtained from DNS ($\ell_m^{\mathrm{DNS}}  = (-\overline{u'w'})^{1/2}\big/(\mathrm{d}\overline{u}/\mathrm{d}z)$, solid lines) with the mixing lengths used in this study. The short dashed lines correspond to the mixing length expression for the shear-dominated sublayer, $\ell_m^{\mathrm{II}}$ (Eq.~\eqref{eq:ml_iner_sl}), while the long-dashed lines correspond to the mixing length formulation in the stratified outer region, $\ell_m^{\mathrm{III}}$ (Eq.~\eqref{eq:mixinglength_sol}). Line colors indicate different cases: green for $Re_{\tau} = 395$, $Ri_{\tau} = 720$; blue for $Re_{\tau} = 550$, $Ri_{\tau} = 720$; and red for $Re_{\tau} = 1000$, $Ri_{\tau} = 600$.}
    \label{fig:ml_compare}
\end{figure}
}
{
Substituting the mixing length expression from Eq.~\eqref{eq:ml_iner_sl} into Eq.~\eqref{eq:reg_ine_sl} yields the velocity-gradient for this region,
}
\begin{equation}\label{eq:reg_ine_sl_g}
    \frac{\mathrm{d} u^+}{\mathrm{d} z^+} = \frac{1}{\kappa z^+}\left(1 + 4.7 \frac{z}{\Lambda}\right),
\end{equation}
which we have shown in figure~\ref{fig:phi_zeta} (see short dash-dotted lines, $z/\Lambda < 1$).

Further, upon substituting $\Lambda$ from Eq.~\eqref{eq:ob_lo_ob} in Eq.~\eqref{eq:reg_ine_sl_g} and integrating, we obtain the velocity profile:
\begin{equation}\label{eq:chn_MO_vel}
    u^+ = \frac{1}{\kappa} \ln(z^+) + \frac{4.7}{\kappa} \frac{2h}{L} \left(1 - \frac{z}{h}\right)^{-1/2} + C_1,
\end{equation}
where $C_1$ is the integration constant, obtained by matching Eq.~\eqref{eq:chn_MO_vel} to the viscous sublayer profile. In neutral conditions, this matching typically occurs at approximately $z^+ \approx 11$ where the viscous and turbulent stresses intersect. With increasing  $Ri_{\tau}$, this point shifts farther away from the wall, as the rate at which turbulent stresses rise is reduced, indicating deeper penetration of buoyancy effects in the near-wall region. This is evident in figure~\ref{fig:stress_schematic}(c), where the blue curves ($Re_{\tau} = 550$) show that the darker shades, representing stronger stratification, rise progressively more slowly. Furthermore, for the same $Ri_{\tau} = 720$, the case at $Re_{\tau} = 395$ (green) shows an even slower rise compared to $Re_{\tau} = 550$. This is because the stronger shear tends to resist the suppression of turbulence by buoyancy forces at higher values of $Re_{\tau}$. The constant used in the present work is described in {Appendix \ref{appA}}.%

{
Interestingly, the half-channel height, $h$, appears in Eq.~\eqref{eq:chn_MO_vel}, indicating the influence of an outer length scale on the mean velocity profile near the wall. In the neutral limit ($L \rightarrow \infty$), however, this dependence vanishes, and the classical logarithmic law is recovered. With the introduction of stratification, a new length scale $\Lambda$ emerges, determined by the local stress distribution. This distribution is constrained by the stress boundary conditions at the wall and at the channel centreline, thereby introducing a dependence on $h$ in the formulation.
}

The {shear-dominated sublayer} extends from $z^+ \approx 30$ to where stratification effects become prominent relative to the turbulent shear. This transition point can be determined analytically by identifying the location where  $z/\Lambda \approx 1$. Substituting $\Lambda = z$ into Eq.~\eqref{eq:ob_lo_ob} yields
\begin{equation}
    \left(\frac{z_t}{h}\right)^3 + \left(\frac{h^2}{L^2} - 3\right)\ \left(\frac{z_t}{h}\right)^2 + 3 \left(\frac{z_t}{h}\right) -1 = 0,
\end{equation}
whose real root gives the normalized transition location, $z_t/h$. 

{
For neutral flows, the transition point $(z_t)$  is located at the channel centerline, indicating the absence of buoyancy effects and the dominance of shear throughout the channel. Consequently, the mixing length given by Eq.~\eqref{eq:ml_iner_sl} is applied up to the channel center. This assumes a logarithmic law of the wall across the remainder of the neutral velocity profile, which is a reasonable approximation given the relatively weak wake contribution observed in channel flows. With increasing stratification, the transition point shifts closer to the wall, marking the onset of the stratified outer region.
}
\subsection{{Stratified outer region} (III)}\label{sec:str_outer_region}

This region begins approximately at the transition point ($z = z_t$), beyond which turbulent stresses keep decreasing while still dominating the stress budget (Figure~\ref{fig:stress_schematic}). Yet, the influence of buoyancy forces becomes increasingly significant $(z/\Lambda > 1)$, with large eddying motion being more effectively suppressed. In this region, Eq.~\eqref{eq:reg_ine_sl} remains valid since turbulent stresses remain much larger than viscous stresses. 

{
However, the mixing length in this region ($\ell_m^{\mathrm{III}}$) differs from that in the shear-dominated sublayer, as stratification effects become significant ($z/\Lambda > 1$). In this case, $\ell_m^{\mathrm{N}}$ is normalized using the Businger relation that is corrected by a factor $(1 - z/h)^{1/2}$. As discussed in \S\ref{sec:most_channel}, similar corrections with varying exponents have been used in previous studies (e.g., \citet{yokoyama1979vertical}, \citet{sorbjan1986similarity}) to extend Eq.~\eqref{eq:obukhovL} across the ABL. Here, the correction is applied directly to the Businger relation and is motivated by the trends observed in figure~\ref{fig:phi_zeta}. Thus,
\begin{equation}\label{eq:mixinglength_sol}
\ell_m^{\mathrm{III}} = \frac{\ell_m^{\mathrm{N}}}{1 + 4.7\zeta}\left(1 - \frac{z}{h}\right)^{-1/2} = \frac{\kappa z}{1 + 4.7\zeta}.
\end{equation}
}

{
This mixing length model is also depicted in figure~\ref{fig:ml_compare} (long dashed lines), agreeing well with the DNS data. The formulation in Eq.~\eqref{eq:mixinglength_sol} reasonably captures the DNS trend in the region dominated by buoyancy effects. 
}

Substituting Eq.~\eqref{eq:mixinglength_sol} in Eq.~\eqref{eq:reg_ine_sl} gives the velocity gradient in the stratified outer region,
\begin{equation}\label{eq:stress_bal3_g}
    \frac{\mathrm{d}u^+}{\mathrm{d}z} = \frac{1}{\kappa z}\left(1 + 4.7 \frac{z}{\Lambda}\right)\left(1 - \frac{z}{h}\right)^{1/2},
\end{equation}
which, after substituting the definition of $\Lambda$ from Eq.~\eqref{eq:ob_lo_ob} and integrating, yields the velocity profile
\begin{equation}\label{eq:str_out_layer}
     u^+ = \frac{1}{\kappa}\ln{\frac{z}{h}}-\frac{4.7}{\kappa} \frac{h}{L} \ln{\left(1 - \frac{z}{h}\right)} + \frac{2}{\kappa} \sqrt{1 - \frac{z}{h}} - \frac{2}{\kappa} \ln{\left(1 + \sqrt{1 - \frac{z}{h}}\right)} + C_2,
\end{equation}
with $C_2$, being an integration constant. Note that a similar expression was derived for open channel flows by \citet{van_de_wiel_cessation_2012, donda_collapse_2015} in the context of ABL.

Furthermore, we observe that $C_2$ approximately corresponds to the centerline velocity of neutral flows at a given $Re_{\tau}$, i.e., $C_2 \approx {1}/{\kappa}\ln{Re_{\tau}} + 5.2$. {This can be reconciled by noting that, in neutral limit ($L \rightarrow \infty$), the shear-dominated sublayer extends up to the centerline. Consequently, the stratified outer region reduces to a single point at $z = h$. In this limit, Eq.~\eqref{eq:str_out_layer} yields the centerline velocity of the neutral flow.} Note also that, in this case, the buoyancy-related term in Eq.~\eqref{eq:str_out_layer} vanishes, and in the overlap region ($0 \ll z \ll h$) the expression also recovers the classical velocity defect law \citep{popeTurbulence}, demonstrating the consistency of the approach.

The profile described by Eq.~\eqref{eq:str_out_layer} results from integrating only the turbulent stress contribution (Eq.~\eqref{eq:reg_ine_sl}). However, farther from the walls, turbulent stresses gradually become comparable to viscous stresses, and the latter must also be taken into account (see also figure~\ref{fig:stress_schematic}). Based on observations across a wide range of cases, this transition typically occurs at a location where $z/\Lambda \approx 8$. { This location also marks the upper bound beyond which $z/\Lambda$ ceases to be a relevant parameter, as the turbulent momentum flux is no longer accurately represented by the total stress in the definition of $\Lambda$.} Beyond this, Eq.~\eqref{eq:reg_ine_sl} is no longer valid, marking the onset of the transition layer. \par
Within the \emph{turbulent--viscous transition layer}, both viscous and turbulent diffusion play comparable roles. {Due to the limited extent of this region, a simple linear blending of the adjacent layers provides a reasonable estimate of the flow velocity here. While this blending approach is not explicitly discussed in the main text, its implementation and resulting composite profile are illustrated in Appendix~{\ref{appE}} and also the accompanying Jupyter notebook.}

\subsection{Viscous core (V)}\label{sec:visc_core}

Finally, towards the center of the channel, stratification may become so influential that the eddies are fully suppressed, turbulence ceases to exist, and the local flow is laminar. Since only viscous stresses contribute to the momentum balance in this region, Eq.~\eqref{eq:nu_t_stress_balance} simplifies and can be integrated, with the centerline velocity ($u_{\mathrm{cl}}^+$) appearing as the constant of integration,

\begin{equation}\label{eq:visc_core}
    \frac{h}{Re_{\tau}} \frac{\mathrm{d}u^+}{\mathrm{d}z} = 1 - \frac{z}{h},
    \quad
    u_{\mathrm{cl}}^+ - u^+ = \frac{Re_{\tau}}{2}\left(1 - \frac{z}{h}\right)^2.
\end{equation}
The distance from the centerline where the viscous and turbulent stresses become comparable can be estimated as the location up to which the mean velocity follows a parabolic profile. A reasonable estimate for this distance, from the channel center, is where
$h/Re_{\tau} \sim 2 \ell_m \sqrt{\tau/\tau_w}$ {(see Appendix \ref{appC})}. This point ($z^l_{\mathrm{V}}$) marks the beginning of the viscous core, which extends up to the centerline. If the flow is strongly stratified, then the turbulent stresses vanish earlier, as can be seen in figure~\ref{fig:stress_schematic}(d). The estimated distance is given as 
\begin{equation}\label{eq:turb_visc2visc_core}
      1 -\frac{z^l_{\mathrm{V}}}{h} =  \frac{1}{\sqrt{Re_L}}\ \sqrt{\frac{4.7}{2\kappa}},
\end{equation}
where $Re_{L} = u_{\tau} L/\nu$ is the friction Reynolds number based on the Obukhov length scale.

\subsection{Important parameterizations for predicting the mean velocity profile}%

The velocity profiles 
(Eqs.~(\ref{eq:chn_MO_vel}, \ref{eq:str_out_layer}, \ref{eq:visc_core}))
depend on the governing input parameters $Re_{\tau}$ and $Ri_{\tau}$. Yet, an important parameter naturally emerging throughout the analysis is a Richardson number based on wall fluxes, $Ri_{\mathrm{w}}=h/L$ (recall Eq.~\ref{eq:obukhovL}). %
Hence, to obtain a closed form of the velocity profiles,  
$Ri_{\mathrm{w}}$ must be estimated a priori. To this end, we derive a relation {(see Appendix \ref{appD})} for it in terms of $Re_{\tau}$ and $Ri_{\tau}$, relying on the Nusselt number scaling proposed by \citet{zonta2022interaction} that was obtained for a range of cases (either simulated at $Pr = 0.71$ or rescaled accordingly).
The proposed scaling is given by:
{
\begin{equation}\label{eq:RihL}
    Ri_{\mathrm{w}} = \frac{h}{L} = \frac{\kappa\ (g/\theta_0)\ q_w}{u_{\tau}^3/h}  \sim \ \left(\frac{Ri_{\tau}}{\sqrt{Re_{\tau}}}\right)^{2/3}.
\end{equation}}

\begin{figure}
    \centering
    \includegraphics[width = \linewidth]{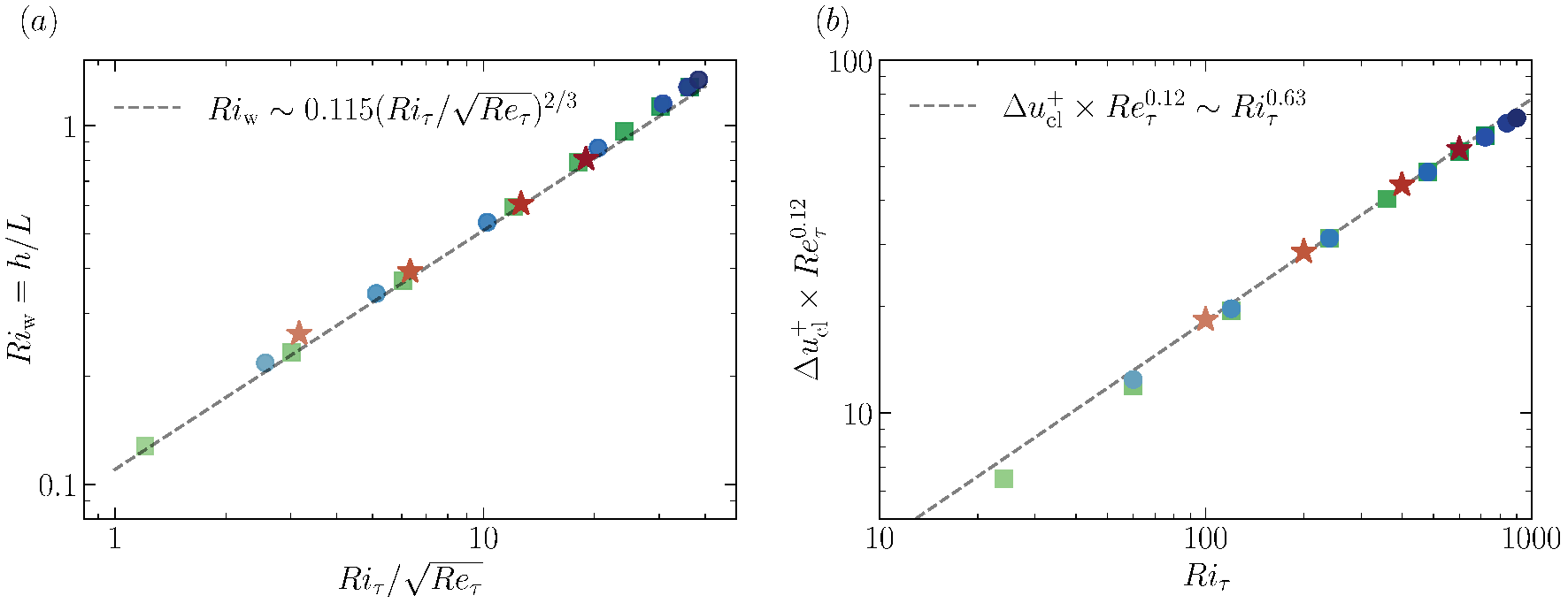}
    \caption{\emph{(a)} The proposed scaling to estimate a priori the parameter $Ri_{\mathrm{w}} = h/L$. \emph{(b)} Parametrization of the deviation of the centerline velocity from the neutral case at same $Re_{\tau}$. Increasing darkness of the marker colors denotes increasing Richardson numbers. The symbols and colors are as indicated in figure~\ref{fig:cases}.}
    \label{fig:parametrize_w}
\end{figure}
Figure~\ref{fig:parametrize_w}(a) shows the agreement of this scaling for the different cases considered in this study. 
Additionally, the viscous core solution requires a closure for stratification effects on the centerline velocity, $u_{\mathrm{cl}}^+$. This is done by modeling its deviation from the neutral case at the same $Re_{\tau}$, given by $u_{\mathrm{cl,N}}^+ \approx {1}/{\kappa} \ln Re_{\tau} + 5.2$.  With stratification, turbulence suppression near the centerline leads to an increased $u_{\mathrm{cl}}^+$ that we found to be {well} captured by:
\begin{equation}
    \Delta u_{\mathrm{cl}}^+ = u_{\mathrm{cl}}^+ - u_{\mathrm{cl,N}}^+ = Ri_{\tau}^{0.63} Re_{\tau}^{-0.12},
\end{equation}
 as shown in figure~\ref{fig:parametrize_w}(b). Indeed, $\Delta u_{\mathrm{cl}}^+$ increases with $Ri_{\tau}$ as expected due to stronger stratification effects, and depends weakly on $Re_{\tau}$, as most of its influence is already accounted in $u_{\mathrm{cl,N}}^+$.

These parametric relations and Eqs.~(\ref{eq:chn_MO_vel},~\ref{eq:str_out_layer},~\ref{eq:visc_core}) are the ingredients needed for reconstructing the mean velocity profile across the entire channel height. In the following section, we compare the predictions against our DNS dataset to evaluate the applicability of MOST in this confined flow configuration.

\vspace{-5pt}

\section{Assessment of the proposed scaling with DNS }
\begin{figure}
    \centering
    \rotatebox{90}{%
        \begin{minipage}{\textheight}
            \centering
            \includegraphics[width=\textwidth]{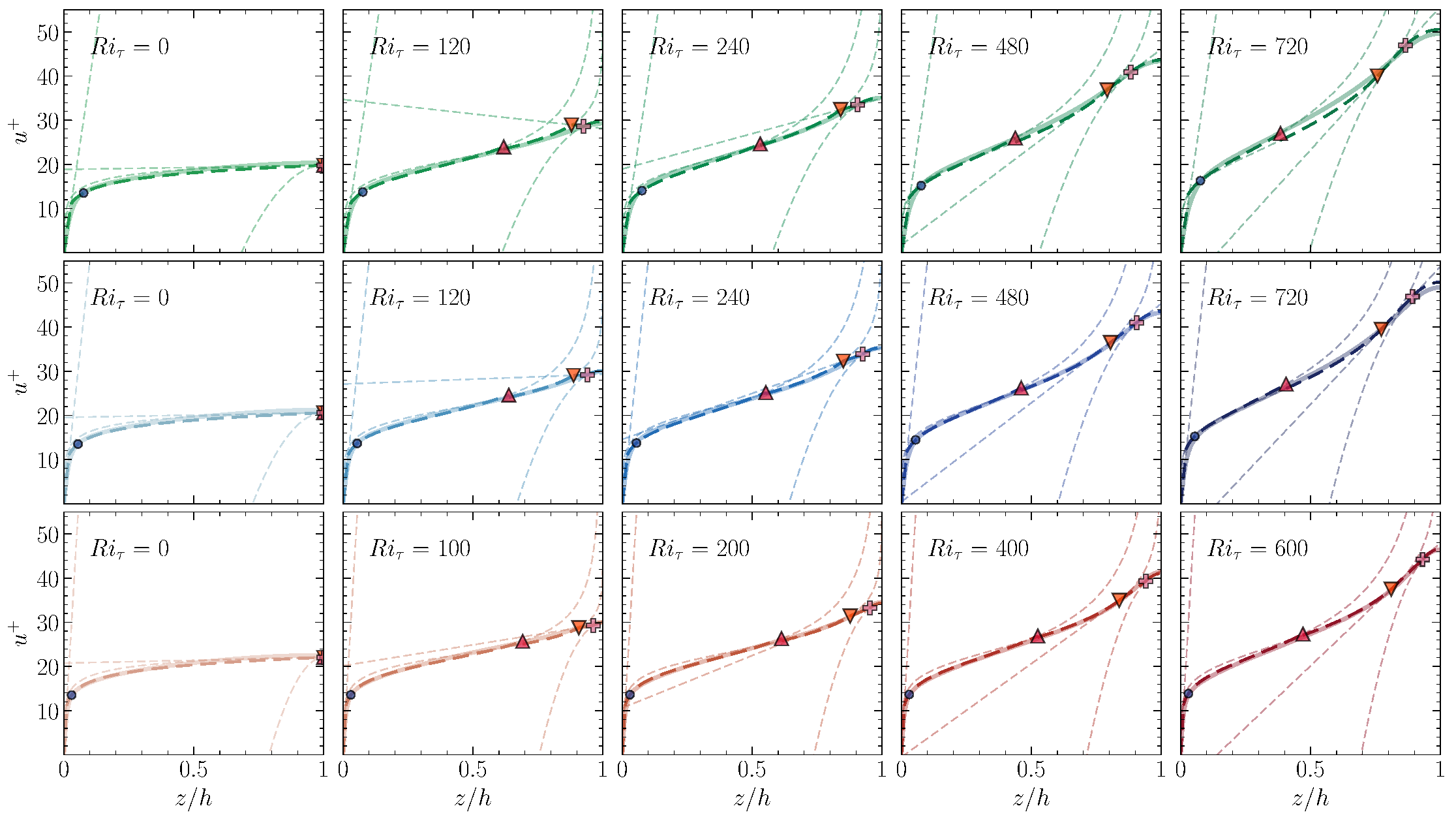}
            \caption{The composite mean velocity profiles obtained from the proposed formulation (dashed lines) are compared with DNS data (solid lines). The markers indicate the boundaries of the different regions, as defined in figure~\ref{fig:stress_schematic}. Results are shown for $Re_{\tau} = 395$ (greens), $Re_{\tau} = 550$ (blues), and $Re_{\tau} = 1000$ (reds). The implementation of the mean velocity profile reconstruction and the estimation of the skin friction coefficient can be found in this \href{https://github.com/sanathbk/Stably-stratified-turbulent-channel-flows}{notebook}.}
            \label{fig:composite}
        \end{minipage}
    }
\end{figure}

{
Figure~\ref{fig:composite} compares the composite mean velocity profile predicted by the proposed approach against DNS results across several $Re_{\tau}$ and $Ri_{\tau}$ cases. The profiles, plotted in linear coordinates from the wall to the channel centreline, provide a global assessment of the model performance. The individual contributions from Eqs.~(\ref{eq:chn_MO_vel}, \ref{eq:str_out_layer}, \ref{eq:visc_core}) are shown as dashed lines, with the highlighted segments indicating the region over which each expression is valid. The symbols are used to distinguish the different flow regions discussed earlier.}

\begin{figure}
    \centering
    \includegraphics[width=\linewidth]{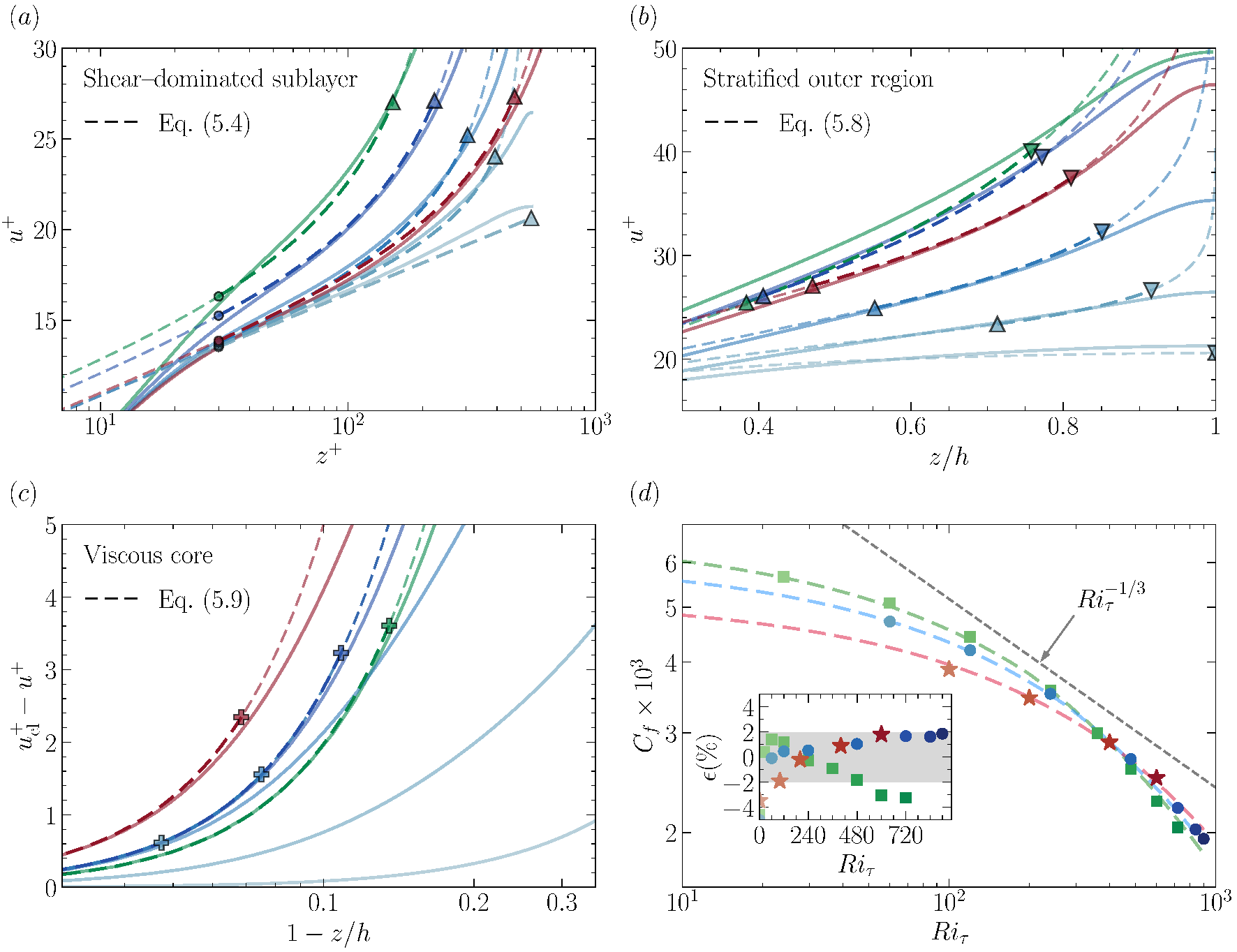}
    \caption{Mean streamwise velocity profiles in stably stratified turbulent channel flows. The solid lines correspond to DNS data which is compared against \emph{(a)} Eq.~\eqref{eq:chn_MO_vel}  in the {shear-dominated sublayer}, \emph{(b)} Eq.~\eqref{eq:str_out_layer}  in the {stratified outer region}, and \emph{(c)} Eq.~\eqref{eq:visc_core}  in the viscous core. The predictions are indicated with dashed lines. Blue curves: $Re_{\tau} = 550$, $Ri_{\tau} = 0,\ 60,\ 240,\ 720$ (increasing darkness). Green: $Re_{\tau} = 395$, $Ri_{\tau} = 720$. Red: $Re_{\tau} = 1000$, $Ri_{\tau} = 600$. $\blacktriangle,\ \blacktriangledown$ and \ding{58} indicate the bounds of different regions. \emph{(d)} Comparison of skin-friction coefficient, $C_f$, predicted by the proposed approach (colored dashed lines) with DNS data (symbols) as a function of $Ri_{\tau}$. The grey dashed line indicates the $C_f \sim Ri_{\tau}^{-1/3}$ scaling shown in \citet{zonta2022interaction}. The inset shows the percentage error ($\epsilon$) between the model and DNS at different $Ri_{\tau}$ across all cases considered. The shaded region indicates a $\pm 2 \%$ error band. The implementation of the mean velocity profile reconstruction and the estimation of the skin friction coefficient can be found in this \href{https://github.com/sanathbk/Stably-stratified-turbulent-channel-flows}{notebook}.}
    \label{fig:all_vel_pro}
\end{figure}

{In addition to this overall comparison, figure~\ref{fig:all_vel_pro} offers a closer picture of the predicted and DNS velocity profiles, focusing on the different flow regions.} To examine the influence of stratification, results are presented for flows with  $Ri_{\tau} = 0, \ 60, \ 240 \ \text{and} \ 720$ at $Re_{\tau} = 550$ (shown in increasingly darker shades of blue). Moreover, to compare the Reynolds number effects, results for flows with $Ri_{\tau} = 720$ are presented at $Re_{\tau} = 395,\ 550$ (green, dark blue). Since the case at $Ri_{\tau} = 720$ was unavailable for $Re_{\tau} = 1000$, the result of the closest available $Ri_{\tau} = 600$ is shown in dark red. \par

In the {shear-dominated sublayer} (II), it can be observed that Eq.~\eqref{eq:chn_MO_vel} accurately predicts the mean velocity profile with discrepancies appearing only for strongly stratified flows; for instance $Ri_{\tau} = 720$ at $Re_{\tau} = 395$ (Figure~\ref{fig:all_vel_pro}a). This is due to insufficient scale separation in the flow, particularly at the imposed stratification. {When $Re_{\tau}$ is not sufficiently large and stratification is strong, the smallest eddies in the flow are not small enough to remain unaffected by buoyancy. As a result, a substantial portion of the turbulence spectrum is suppressed.} The estimated extent of the {shear-dominated sublayer} ($z_t$) is indicated for different cases with $\blacktriangle$. If $z_t^+ < 150$, then the buoyancy effects are perceived too close to the wall, disrupting the near-wall region of the flow and thus violating the assumption of {a logarithmic velocity profile in this region. }
\par

In the {stratified outer region} (III), predictions with  Eq.~\eqref{eq:str_out_layer}, with $C_2 \approx 1/\kappa \ln{Re_{\tau}} + 5.2$, agree well with all the cases except for the case at $Re_{\tau} = 395$ (Figure~\ref{fig:all_vel_pro}b), for the same reason as described above. While the equation effectively reproduces the general shape of the velocity profile, we observe a shift for low $Re_{\tau}$ cases.
The point marked with $\blacktriangledown$ ($z_{III}^u$) in figure~\ref{fig:all_vel_pro}(b) indicates the end of the {stratified outer region} and the onset of the turbulent-viscous transition layer, where the influence of viscous stresses becomes comparable to that of turbulent stresses.
Finally, Figure~\ref{fig:all_vel_pro}(c) also confirms the agreement of the parabolic velocity profile with DNS data in the viscous core (V). {The velocity profile in this region follows a parabolic form parametrized solely by $Re_{\tau}$. Hence, a distinct parabolic curve is obtained for each $Re_{\tau} = 395,\ 550,\ 1000$. The extent to which the DNS profiles follow the parabolic form, measured from the channel center, increases with stratification. This is evident for $Ri_{\tau} = 0,\ 60,\ 240,\ 720$ (increasing shades of blue) at  $Re_{\tau} = 550$. Specifically, the $Ri_{\tau} = 720$ case follows the parabolic profile over a larger portion of the core compared to $Ri_{\tau} = 60\ \text{and}\ 240$, while $Ri_{\tau} = 0$ does not exhibit this behavior in the center. The boundary marking the approximate location beyond which the velocity profiles begin to diverge from the parabola is indicated by the symbol \ding{58}. We observe that for $Ri_{\tau} =720$ at $Re_{\tau} = 395,\ 550$ and $Ri_{\tau} = 600$ at $Re_{\tau} = 1000$, this is estimated reasonably well. However, at lower $Ri_{\tau}$ values, corresponding to $Re_{\tau} = 550$, this boundary is estimated to be at a slightly larger distance from the channel center than the actual location where the velocity profiles begin to diverge from the parabola. Nevertheless, this difference is visually exaggerated due to the logarithmic horizontal axis, and the estimate is acceptable for practical purposes.}

The mean velocity profile obtained from the proposed approach can be integrated to obtain bulk velocity ($u_b$), and consequently estimate the skin friction coefficient ($C_f = 2\tau_w/\rho u_b^2 = 2/{u_b^+}^2$). Figure~\ref{fig:all_vel_pro}(d) compares these estimates with values obtained from DNS, with the scaling $C_f \sim Ri_{\tau}^{-1/3}$ reported by \citet{zonta2022interaction} also shown for reference. While this relation may hold in a narrower parameter range, it tends to deviate from DNS results over the wider parameter space explored in this study. Integrating the velocity profile obtained from the proposed approach yields consistently accurate results, illustrating the robustness of the present framework. The relative errors, defined as $\epsilon = (C_f^{\mathrm{DNS}} - C_f)/C_f^{\mathrm{DNS}}$, are generally within a $2\%$ margin, with the exception of the high $Ri_{\tau}$ cases at the lowest $Re_{\tau} = 395$, where deviations up to $4\%$ are observed. This slightly larger error is, once more, an expected consequence of the lack of scale separation in these cases.

%% file: conclusion.tex
In this work, we have leveraged a framework derived for atmospheric science -- MOST -- and adopted it to describe the mean velocity in a stably stratified turbulent internal (channel) flow. The linear Businger relation is employed as the stability correction function to characterize the effect of stable stratification on the mean velocity profile.
Based on the relative contributions of viscous and turbulent stresses and the local stability parameter ($z/\Lambda$), distinct flow regions were identified.
The classical viscous sublayer is followed by an \emph{{shear-dominated sublayer}}, where the mean flow is properly characterized by wall values and stratification effects are minimal. This is followed by a \emph{{stratified outer region}} extending from $z\approx\Lambda(z)$ %
to $z\approx8\Lambda(z)$. 
Turbulent stresses then become comparable to the viscous stresses in the \emph{turbulent-viscous transition layer}. Finally, the flow becomes laminar in the \emph{viscous core}, whose onset was estimated to scale inversely with ($Re_{L}^{-1/2}$). 
In the analysis, the wall Richardson number, $Ri_{\mathrm{w}} = h/L$ and deviation from the neutral centreline velocity $\Delta u_\mathrm{cl}^+$, emerge as an important parameters. Accordingly, we have proposed parametric relations to estimate them that enable an accurate reconstruction of the mean velocity profile for a wide range of governing parameters $Re_{\tau}$ and $Ri_{\tau}$. Our findings highlight MOST's robustness by demonstrating that a theory originally developed and validated for atmospheric flows at scales of ${\sim}100\,\mathrm{m}$ effectively extends to stably stratified turbulence in internal flows, applicable in systems at scales orders of magnitude smaller (e.g., ${\sim}0.01\,\mathrm{m}$). {The framework effectively captures stratification effects despite relying on empirical, non-universal stability correction functions.}
Crucially, the reconstructed profiles enable a reliable estimation of the skin friction coefficient, with errors generally within $3\%$. This provides a framework for quantifying pressure losses, offering practical value for modeling and designing flow systems in engineering applications where stratification is important.

\par

\textbf{Declaration of Interests.} The authors report no conflict of interest.

\textbf{Acknowledgements.} We thank Baptiste Hardy for valuable discussions on MOST and stably stratified turbulent boundary layers. We are also grateful for the fruitful exchanges with Giandomenico Lupo and Asif Hasan. We further thank Dr. Francesco Zonta for providing DNS data. This work was supported by the European Research Council (grant no. ERC-2019-CoG-864660, Critical), and NVIDIA Corporation (Academic Grant \emph{Turbulent forced convection beyond the Oberbeck-Boussinesq hypothesis}). 

%% file: appendix.tex
\section{Velocity profile in shear-dominated sublayer}\label{appA}
In this section, we elaborate on the shear-dominated sublayer (introduced in \S\ref{sec:iner_sl} in the main text), with particular focus on estimating the constant of integration in the velocity profile expression. 

In the shear-dominated sublayer, the total stress is dominated by turbulent shear stress, which can be approximated by the wall shear stress. This forms the basis for deriving the mean velocity profile in this region, as outlined below,
\begin{equation}
    \ell_m \frac{\mathrm{d} \overline u}{\mathrm{d} z} = u_{\tau}\left(1 - \frac{z}{h}\right)^{1/2},
\end{equation}
where $\ell_m$ in region $\mathrm{II}$ is formulated using $\ell_m^{\mathrm{N}}$ (Eq.~\eqref{eq:ml_neutral}), and the Businger relation as,
\begin{equation}
    \ell_m^{\mathrm{II}} = \frac{\ell_m^{\mathrm{N}}}{\phi} =\frac{\ell_m^{\mathrm{N}}}{1 + 4.7 \ \zeta} = \frac{\kappa z}{1 + 4.7 \ \zeta}\left(1 - \frac{z}{h}\right)^{1/2}\text{;}
    \quad \ \ \
    \zeta = \frac{z}{\Lambda}.
\end{equation}
We also know that,
\begin{equation}\label{eq:localOb}
    \Lambda = L \left(1 - \frac{z}{h}\right)^{3/2}.
\end{equation}
Using Eq.~\eqref{eq:localOb} and integrating the gradient yields the velocity profile in the shear-dominated sublayer,

\begin{equation}\label{eq:app_sl}
    u^+ = \frac{1}{\kappa}\ln{z^+} + \frac{4.7}{\kappa} \frac{2h}{L} \left(1 - \frac{z}{h}\right)^{-1/2} + C_0,
\end{equation}
where $C_0$ is a constant.
As indicated in the main text, the constant is determined by matching the profiles between the viscous and shear-dominated sublayers. For instance, if the linear profile, $u^+ = z^+$, is matched with Eq.~\eqref{eq:app_sl} at $z^+ = {z_\mathrm{M}^+}$ then the constant $C_0$ is given as,
\begin{equation}
    C_0 = {z_\mathrm{M}^+} - \frac{1}{\kappa}\ln({z_\mathrm{M}^+}) - \frac{4.7}{\kappa} \frac{2h}{L} - \frac{4.7}{\kappa} \frac{h}{L}\frac{{z_\mathrm{M}^+}}{Re_{\tau}},
\end{equation}
where the position ${z_\mathrm{M}^+}$ is defined as;
\begin{equation}
    {z_\mathrm{M}^+} = 11.1 + C_z \ \mathrm{max}\left(0, \frac{Ri_{\tau}}{Re_{\tau}} - 0.5\right),
\end{equation}
with $C_z = 2$ used in the present work.\par

In neutral flows, the matching occurs at $z_\mathrm{M}^+ = 11.1$.  For stably stratified flows, the constant $C_0$ is determined at $z_\mathrm{M}^+ = 11.1$ for $Ri_{\tau}/{Re_{\tau}} < 0.5$, beyond which the intersection point was observed to scale linearly with stratification strength. The value of $0.5$ is based on observations. The slope, $C_z$, may depend on $Re_{\tau}$, but precise determination of the function would require a significantly larger dataset. While other functional forms of $z_\mathrm{M}^+$ could also be used to fit the data, the primary focus here is on the conceptual framework and formulation rather than a detailed empirical fit.

\section{Extent of the viscous core}\label{appC}
This section provides additional details on the estimation of the transition point  $z^l_{\mathrm{V}}$ discussed in \S\ref{sec:visc_core} \emph{(Viscous core (V))} of the main manuscript.

The mean shear can be evaluated as the root of the quadratic streamwise momentum balance equation, %

\begin{equation}\label{eq:stress_bal4}
    \frac{\mathrm{d}u^+}{\mathrm{d}z}  = \frac{2 \tau/\tau_w}{\cfrac{h}{Re_{\tau}} + \sqrt{\cfrac{h^2}{Re_{\tau}^2} + 4l_m^2\cfrac{\tau}{\tau_w}}}.
\end{equation}
In the viscous core, turbulent stresses vanish, and viscous diffusion is the only mechanism of momentum transport, implying that $4l_m^2{\tau}/{\tau_w} = 0$, in the above equation. However, moving away from the channel centerline, viscous stresses gradually give in to turbulent stresses. The location where these two contributions become comparable marks the point at which turbulent stresses start to dominate the flow dynamics:

\begin{equation}\label{eq:ex_vi}
    \frac{h}{Re_{\tau}} \sim 2 l_m \sqrt\frac{\tau}{\tau_w}
\quad \text{where,} \ \ \ 
    l_m = {\kappa z} \left(1 + 4.7\frac{z}{\Lambda}\right)^{-1}.
\end{equation}
In the limit, $z\rightarrow h$, $\Lambda \rightarrow 0$, therefore,
\begin{equation}
    l_m \sim \frac{\kappa \Lambda}{4.7}\sim \frac{\kappa L}{4.7}\left(1 - \frac{z}{h}\right)^{3/2}.
\end{equation}
Using this in Eq.~\eqref{eq:ex_vi},
\begin{equation}
    \frac{h}{Re_{\tau}} \sim 2\frac{\kappa L}{4.7}\left(1 - \frac{z}{h}\right)^{2}\mathrm{.}
\end{equation}
The wall-normal distance that marks the extent of the viscous core is denoted by ${z^l_{\mathrm{V}}}$,
\begin{equation}
    1 -\frac{z^l_{\mathrm{V}}}{h} \sim \sqrt{\frac{4.7}{2 \kappa}} \sqrt{\frac{h}{L}} \sqrt{\frac{1}{Re_{\tau}}}\mathrm{.}
\end{equation}

\section{Parametrization of wall Richardson number ($Ri_\mathrm{w} = h/L$)}\label{appD}
This section outlines the step-by-step derivation of the parametric relation used in the main text to express the Obukhov length scale in terms of $Re_{\tau}$ and $Ri_{\tau}$ (Eq.~\eqref{eq:RihL} in the main text). Consider the definition of the Obukhov length scale,

\begin{equation}
    L = \frac{u_{\tau}^{3}}{\kappa (g/\theta_0) q_w};
    \quad 
    \text{where,}\
    q_w = \alpha \left.\frac{\mathrm{d}{\overline{\theta}}}{\mathrm{d}z}\right|_w = \frac{\nu}{Pr} \left.\frac{\mathrm{d}{\overline \theta}}{\mathrm{d}z}\right|_w.
\end{equation}

\begin{equation}
    \implies \frac{h}{L} = h\ \frac{\kappa (g/\theta_0) q_w}{u_{\tau}^{3}} = \kappa \ h\ \frac{g}{\theta_0}\ \frac{ 1}{u_{\tau}^{3}}\ \frac{\nu}{Pr} \left.\frac{\mathrm{d}{\overline \theta}}{\mathrm{d}z}\right|_w,
\end{equation}
where $Pr$ is the Prandtl number of the flow. Multiplying and dividing by $h$ and $\Delta \rho/\rho_0$, and grouping terms corresponding to $Re_{\tau}$ and $Ri_{\tau}$;

\begin{equation}
    \frac{h}{L} \sim \frac{h}{\theta_0}\ \frac{\rho_0}{\Delta \rho} \ \frac{ \Delta \rho g h}{\rho_0 u_{\tau}^2} \ \frac{\nu}{u_{\tau}h}\frac{1}{Pr}\ \left.\frac{\mathrm{d}{\theta}}{\mathrm{d}z}\right|_w. 
\end{equation}
We also use the ideal gas relation, $\Delta \rho/\rho_0 = \Delta \theta/\theta_0$,
\begin{equation}
    \frac{h}{L} \sim \ \frac{1}{Pr}\frac{Ri_{\tau}}{Re_{\tau}}\  \frac{h}{\Delta \theta} \  \left.\frac{\mathrm{d}{\theta}}{\mathrm{d}z}\right|_w.  
\end{equation}

From \citet{zonta2022interaction},

\begin{equation}
    Nu \sim \left(\frac{Re_{\tau}^2}{Ri_{\tau}}\right)^{1/3};
    \quad
    \text{where,}\ \ \
    Nu = \frac{2q_w h}{\lambda \Delta \theta} \sim \frac{h}{\Delta \theta} \  \left.\frac{\mathrm{d}{\theta}}{\mathrm{d}z}\right|_w.  
\end{equation}

Substituting this in the $h/L$ relation\footnote{{
Note that $Ri_w = h/L$ is termed a Richardson number because it can be expressed in a conventional form that denotes the relative importance of turbulence destruction by buoyancy forces and turbulence production.%
}},

\begin{equation}
    Ri_\mathrm{w} = \frac{\kappa\ (g/\theta_0)\ q_w}{u_{\tau}^3/h} = \frac{h}{L} \sim \ \frac{1}{Pr}\left(\frac{Ri_{\tau}}{\sqrt{Re_{\tau}}}\right)^{2/3}.
\end{equation}

It is important to note that, although the Prandtl number ($Pr$) appears in this analysis, its explicit role in the derivation of Zonta et al.'s relation remains unclear. In this study, all of the results are presented for constant $Pr = 0.71$. Moreover, the applicability of MOST across a broad range of $Pr$ values has not been thoroughly examined and requires a dedicated investigation beyond the scope of this work.

\section{Composite mean velocity profile}\label{appE}
The mean velocity profiles in the viscous sublayer, shear-dominated sublayer, stratified outer region, and the viscous core are described in Eqs.~(\ref{eq:chn_MO_vel},~\ref{eq:str_out_layer},~\ref{eq:visc_core}), respectively. To obtain the composite mean velocity profile across the entire channel, the velocity in the turbulent–viscous transition layer must be determined. This is achieved by smoothly blending the velocity profiles from the adjacent regions. If $u_{\mathrm{III}}$ and $u_{\mathrm{V}}$ denote the predicted velocities in regions $\mathrm{III}$ and $\mathrm{V}$, then the velocity profile in region $\mathrm{IV}$ is,
\begin{equation}
    u_{\mathrm{IV}} = \gamma\ u_{\mathrm{III}} + (1-\gamma)\ u_{\mathrm{V}};
    \quad
    \text{where,}
    \quad
    \gamma = \frac{z^l_{\mathrm{V}} - z}{z^l_{\mathrm{V}} - z^u_{\mathrm{III}}}.
\end{equation}
$z^u_{\mathrm{III}}$ and $z^l_{\mathrm{V}}$ are the upper and lower bounds of regions $\mathrm{III}$ and $\mathrm{V}$ respectively.
\par

%% file: JFM-template.bbl
\begin{thebibliography}{35}
\expandafter\ifx\csname natexlab\endcsname\relax\def\natexlab#1{#1}\fi
\def\au#1{#1} \def\ed#1{#1} \def\yr#1{#1}\def\at#1{#1}\def\jt#1{\textit{#1}} \def\bt#1{#1}\def\bvol#1{\textbf{#1}} \def\vol#1{#1} \def\pg#1{#1} \def\publ#1{#1}\def\arxiv#1{#1}\def\org#1{#1}\def\st#1{\textit{#1}}

\bibitem[Armenio \& Sarkar(2002)]{armenio2002investigation}
{\sc \au{Armenio, V.} \& \au{Sarkar, S.}} \yr{2002}  \at{An investigation of stably stratified turbulent channel flow using large-eddy simulation}.  \jt{J. Fluid Mech.}  \bvol{459},  \pg{1--42}.

\bibitem[Beljaars \& Holtslag(1991)]{beljaars1991flux}
{\sc \au{Beljaars, A.C.M.} \& \au{Holtslag, A.A.M.}} \yr{1991}  \at{Flux parameterization over land surfaces for atmospheric models}.  \jt{J. Appl. Meteorol. Climatol.}  \bvol{30}~(3),  \pg{327--341}.

\bibitem[Brethouwer {\em et~al.\/}(2007)Brethouwer, Billant, Lindborg \& Chomaz]{brethouwer_scaling_2007}
{\sc \au{Brethouwer, G.}, \au{Billant, P.}, \au{Lindborg, E.} \& \au{Chomaz, J.-M.}} \yr{2007}  \at{Scaling analysis and simulation of strongly stratified turbulent flows}.  \jt{J. Fluid Mech.}  \bvol{585},  \pg{343--368}.

\bibitem[Businger \& Yaglom(1971)]{businger1971introduction}
{\sc \au{Businger, J.} \& \au{Yaglom, A.}} \yr{1971}  \at{Introduction to {Obukhov}'s paper on ‘turbulence in an atmosphere with a non-uniform temperature’}.  \jt{Bound.-Layer Meteorol.}  \bvol{2},  \pg{3--6}.

\bibitem[Caulfield(2021)]{caulfield_layering_2021}
{\sc \au{Caulfield, C.P.}} \yr{2021}  \at{Layering, instabilities, and mixing in turbulent stratified flows}.  \jt{Annu. Rev. Fluid Mech.}  \bvol{53}~(1),  \pg{113--145}.

\bibitem[Chenge \& Brutsaert(2005)]{chenge2005flux}
{\sc \au{Chenge, Y.} \& \au{Brutsaert, W.}} \yr{2005}  \at{Flux-profile relationships for wind speed and temperature in the stable atmospheric boundary layer}.  \jt{Bound.-Layer Meteorol.}  \bvol{114}~(3),  \pg{519--538}.

\bibitem[Cook \& Riley(1996)]{COOK1996263}
{\sc \au{Cook, A.W.} \& \au{Riley, J.J.}} \yr{1996}  \at{Direct numerical simulation of a turbulent reactive plume on a parallel computer}.  \jt{J. Comput. Phys.}  \bvol{129}~(2),  \pg{263--283}.

\bibitem[Costa(2018)]{COSTA20181853}
{\sc \au{Costa, P.}} \yr{2018}  \at{A {FFT}-based finite-difference solver for massively-parallel direct numerical simulations of turbulent flows}.  \jt{Comput. Math. Appl.}  \bvol{76}~(8),  \pg{1853--1862}.

\bibitem[Donda {\em et~al.\/}(2015)Donda, {van Hooijdonk}, Moene, Jonker, {van Heijst}, Clercx \& {van de Wiel}]{donda_collapse_2015}
{\sc \au{Donda, J.M.M.}, \au{{van Hooijdonk}, I.G.S.}, \au{Moene, A.F.}, \au{Jonker, H.J.J.}, \au{{van Heijst}, G.J.F.}, \au{Clercx, H.J.H.} \& \au{{van de Wiel}, B.J.H.}} \yr{2015}  \at{Collapse of turbulence in stably stratified channel flow: a transient phenomenon}.  \jt{Q. J. R. Meteorol. Soc.}  \bvol{141}~(691),  \pg{2137--2147}.

\bibitem[Dyer(1974)]{dyer1974review}
{\sc \au{Dyer, A.J.}} \yr{1974}  \at{A review of flux-profile relationships}.  \jt{Bound.-Layer Meteorol.}  \bvol{7}~(3),  \pg{363--372}.

\bibitem[Fukui {\em et~al.\/}(1983)Fukui, Nakajima \& Ueda]{fukui_laboratory_1983}
{\sc \au{Fukui, K.}, \au{Nakajima, M.} \& \au{Ueda, H.}} \yr{1983}  \at{A laboratory experiment on momentum and heat transfer in the stratified surface layer}.  \jt{Q. J. R. Meteorol. Soc.}  \bvol{109}~(461),  \pg{661--676}.

\bibitem[Garcia-Villalba \& {Del Alamo}(2011)]{garcia2011turbulence}
{\sc \au{Garcia-Villalba, M.} \& \au{{Del Alamo}, J.C.}} \yr{2011}  \at{Turbulence modification by stable stratification in channel flow}.  \jt{Phys. Fluids}  \bvol{23}~(4),  \pg{045104}.

\bibitem[Garg {\em et~al.\/}(2000)Garg, Ferziger, Monismith \& Koseff]{garg2000stably}
{\sc \au{Garg, R.P.}, \au{Ferziger, J.H.}, \au{Monismith, S.G.} \& \au{Koseff, J.R.}} \yr{2000}  \at{Stably stratified turbulent channel flows. i. stratification regimes and turbulence suppression mechanism}.  \jt{Phys. Fluids}  \bvol{12}~(10),  \pg{2569--2594}.

\bibitem[Grachev {\em et~al.\/}(2015)Grachev, Andreas, Fairall, Guest \& Persson]{grachev_similarity_2015}
{\sc \au{Grachev, A.A.}, \au{Andreas, E.L.}, \au{Fairall, C.W.}, \au{Guest, P.S.} \& \au{Persson, P.O.G.}} \yr{2015}  \at{Similarity theory based on the {Dougherty}–{Ozmidov} length scale}.  \jt{Q. J. R. Meteorol. Soc.}  \bvol{141}~(690),  \pg{1845--1856}.

\bibitem[Gryning {\em et~al.\/}(2007)Gryning, Batchvarova, Brümmer, Jørgensen \& Larsen]{gryning2007extension}
{\sc \au{Gryning, S.-E.}, \au{Batchvarova, E.}, \au{Brümmer, B.}, \au{Jørgensen, H.} \& \au{Larsen, S.}} \yr{2007}  \at{On the extension of the wind profile over homogeneous terrain beyond the surface boundary layer}.  \jt{Bound.-Layer Meteorol.}  \bvol{124}~(2),  \pg{251--268}.

\bibitem[Holtslag \& Nieuwstadt(1986)]{holtslag1986scaling}
{\sc \au{Holtslag, A.A.M.} \& \au{Nieuwstadt, F.T.M.}} \yr{1986}  \at{Scaling the atmospheric boundary layer}.  \jt{Bound.-Layer Meteorol.}  \bvol{36}~(1),  \pg{201--209}.

\bibitem[Iida {\em et~al.\/}(2002)Iida, Kasagi \& Nagano]{iida2002direct}
{\sc \au{Iida, O.}, \au{Kasagi, N.} \& \au{Nagano, Y.}} \yr{2002}  \at{Direct numerical simulation of turbulent channel flow under stable density stratification}.  \jt{Int. J. Heat Mass Transf.}  \bvol{45}~(8),  \pg{1693--1703}.

\bibitem[Li {\em et~al.\/}(2016)Li, Salesky \& Banerjee]{li_connections_2016}
{\sc \au{Li, D.}, \au{Salesky, S.T.} \& \au{Banerjee, T.}} \yr{2016}  \at{Connections between the {Ozmidov} scale and mean velocity profile in stably stratified atmospheric surface layers}.  \jt{J. Fluid Mech.}  \bvol{797},  \pg{R3}.

\bibitem[Majda \& Sethian(1985)]{MAJDA1985}
{\sc \au{Majda, A.} \& \au{Sethian, J.}} \yr{1985}  \at{The derivation and numerical solution of the equations for zero mach number combustion}.  \jt{Combust. Sci. Technol.}  \bvol{42}~(3-4),  \pg{185--205}.

\bibitem[Moestam \& Davidson(2005)]{moestam2005numerical}
{\sc \au{Moestam, R.} \& \au{Davidson, L.}} \yr{2005}  \at{Numerical simulations of a thermocline in a pressure-driven flow between two infinite horizontal plates}.  \jt{Phys. Fluids}  \bvol{17}~(7),  \pg{075109}.

\bibitem[Monin \& Obukhov(1954)]{monin1954basic}
{\sc \au{Monin, A.} \& \au{Obukhov, A.}} \yr{1954}  \at{Basic laws of turbulent mixing in the surface layer of the atmosphere}  \pg{pp. 163--187}.

\bibitem[Nieuwstadt(1984)]{nieuwstadt1984turbulent}
{\sc \au{Nieuwstadt, F.T.M.}} \yr{1984}  \at{The turbulent structure of the stable, nocturnal boundary layer}.  \jt{J. Atmos. Sci.}  \bvol{41}~(14),  \pg{2202--2216}.

\bibitem[Nieuwstadt(2005)]{nieuwstadt2005direct}
{\sc \au{Nieuwstadt, F.T.M.}} \yr{2005}  \at{Direct numerical simulation of stable channel flow at large stability}.  \jt{Bound.-Layer Meteorol.}  \bvol{116}~(2),  \pg{277--299}.

\bibitem[Obukhov(1971)]{obukhov1971turbulence}
{\sc \au{Obukhov, A.}} \yr{1971}  \at{Turbulence in an atmosphere with a non-uniform temperature}.  \jt{Bound.-Layer Meteorol.}  \bvol{2}~(1),  \pg{7--29}.

\bibitem[Pirozzoli(2014)]{pirozzoli_revisiting_2014}
{\sc \au{Pirozzoli, S.}} \yr{2014}  \at{Revisiting the mixing-length hypothesis in the outer part of turbulent wall layers: mean flow and wall friction}.  \jt{J. Fluid Mech.}  \bvol{745},  \pg{378--397}.

\bibitem[Pope(2000)]{popeTurbulence}
{\sc \au{Pope, S.B.}} \yr{2000} {\em Turbulent Flows\/}.  \publ{Cambridge University Press}.

\bibitem[Salesky {\em et~al.\/}(2013)Salesky, Katul \& Chamecki]{salesky_buoyancy_2013}
{\sc \au{Salesky, S.T.}, \au{Katul, G.G.} \& \au{Chamecki, M.}} \yr{2013}  \at{Buoyancy effects on the integral lengthscales and mean velocity profile in atmospheric surface layer flows}.  \jt{Phys. Fluids}  \bvol{25}~(10),  \pg{105101}.

\bibitem[Shen {\em et~al.\/}(2025)Shen, Liu, Lu \& Stevens]{Shen_Liu_Lu_Stevens_2025}
{\sc \au{Shen, Zhouxing}, \au{Liu, Luoqin}, \au{Lu, Xi-Yun} \& \au{Stevens, Richard~J.A.M.}} \yr{2025}  \at{Mean turbulent momentum fluxes and wind deficits in nocturnal stable atmospheric boundary layers}.  \jt{J. Fluid Mech.}  \bvol{1017},  \pg{A5}.

\bibitem[Sorbjan(1986)]{sorbjan1986similarity}
{\sc \au{Sorbjan, Z.}} \yr{1986}  \at{On similarity in the atmospheric boundary layer}.  \jt{Bound.-Layer Meteorol.}  \bvol{34}~(4),  \pg{377--397}.

\bibitem[Stiperski \& Calaf(2023)]{stiperski_generalizing_2023}
{\sc \au{Stiperski, I.} \& \au{Calaf, M.}} \yr{2023}  \at{Generalizing {Monin}--{Obukhov} similarity theory (1954) for complex atmospheric turbulence}.  \jt{Phys. Rev. Lett.}  \bvol{130}~(12),  \pg{124001}.

\bibitem[{van de Wiel} {\em et~al.\/}(2012){van de Wiel}, Moene \& Jonker]{van_de_wiel_cessation_2012}
{\sc \au{{van de Wiel}, B.J.H.}, \au{Moene, A.F.} \& \au{Jonker, H.J.J.}} \yr{2012}  \at{The cessation of continuous turbulence as precursor of the very stable nocturnal boundary layer}.  \jt{J. Atmos. Sci.}  \bvol{69}~(11),  \pg{3097--3115}.

\bibitem[Webb(1970)]{webb1970profile}
{\sc \au{Webb, E.K.}} \yr{1970}  \at{Profile relationships: The log-linear range, and extension to strong stability}.  \jt{Q. J. R. Meteorol. Soc.}  \bvol{96}~(407),  \pg{67--90}.

\bibitem[Yokoyama {\em et~al.\/}(1979)Yokoyama, Gamo \& Yamamoto]{yokoyama1979vertical}
{\sc \au{Yokoyama, O.}, \au{Gamo, M.} \& \au{Yamamoto, S.}} \yr{1979}  \at{The vertical profiles of the turbulence quantities in the atmospheric boundary layer}.  \jt{J. Meteorol. Soc. Japan. Ser. II}  \bvol{57}~(3),  \pg{264--272}.

\bibitem[Zonta {\em et~al.\/}(2022)Zonta, Sichani \& Soldati]{zonta2022interaction}
{\sc \au{Zonta, F.}, \au{Sichani, P.H.} \& \au{Soldati, A.}} \yr{2022}  \at{Interaction between thermal stratification and turbulence in channel flow}.  \jt{J. Fluid Mech.}  \bvol{945},  \pg{A3}.

\bibitem[Zonta \& Soldati(2018)]{zonta_stably_2018}
{\sc \au{Zonta, F.} \& \au{Soldati, A.}} \yr{2018}  \at{Stably stratified wall-bounded turbulence}.  \jt{Appl. Mech. Rev.}  \bvol{70}~(4),  \pg{040801}.

\end{thebibliography}
